\documentclass{article}
\usepackage{spconf,amsmath,graphicx}

\title{Requirements for Mass Adoption of Assistive Listening Technology by the General Public}
\name{Thomas B. Kaufmann \qquad Mehdi Foroogozar \qquad Julie Liss \qquad Visar Berisha}
\address{Arizona State University\\Department of Speech and Hearing Science\\Tempe, AZ, USA}

\begin{document}
\maketitle

\begin{abstract}
Assistive listening systems (ALSs) dramatically increase speech intelligibility and reduce listening effort. It is very likely that essentially everyone, not only individuals with hearing loss, would benefit from the increased signal-to-noise ratio an ALS provides in almost any listening scenario. However, ALSs are rarely used by anyone other than people with severe to profound hearing losses. To date, the reasons for this poor adoption have not been systematically investigated.

The authors hypothesize that the reasons for poor adoption of assistive listening technology include (1) an inability to use personally owned receiving devices, (2) a lack of high-fidelity stereo sound, (3) receiving devices not providing an unoccluded listening experience, (4) distortion from alignment delay and (5) a lack of automatic connectivity to an available assistive listening audio signal.

We propose solutions to each of these problems in an effort to pave the way for mass adoption of assistive listening technology by the general public.
\end{abstract}


\section{Introduction}
\label{sec:intro}

Large auditoriums and performance spaces with more than 100 seats are typically equipped with a sound reinforcement system, to provide adequate sound pressure levels to each seat in the audience, allowing listeners to hear music clearly and comprehend speech with relative ease \cite{Nabelek1986ComparisonAuditorium}. However, speech comprehension in a noisy and reverberating space, such as a large auditorium, classroom, theater, or sports stadium becomes significantly more challenging, particularly for individuals with hearing loss, including hearing aid users and cochlear implant users \cite{Kim2014AInstruments,Dorman2017SpeechImplants}. To improve listening comprehension in such settings, assistive listening systems (ALSs) exist. While ALSs are frequently used by individuals with hearing loss, they are typically not used by the general public.

An ALS creates a wireless connection between a venue's sound system and a receiving device that a listener in the audience wears, bypassing the acoustic space. ALS have been shown to increase the signal-to-noise ratio for the listener by as much as 15 to 20\,dB, vastly outperforming noise-reduction algorithms and directional microphones \cite{hawkins_comparisons_1984, lewis_speech_2004}. Available audio transmission modalities that are compliant with many international accessibility laws include radio frequency (RF), infrared (IR), and magnetic induction \cite{Kaufmann2015HearingTechnology}. While RF and IR systems require each individual listener to borrow a receiving device from the venue, audio frequency induction loop systems (AFILS) allow individuals with telecoil-equipped hearing aids and cochlear implants to connect directly to the venue's audio broadcast \cite{Kaufmann2015HearingTechnology,Riehle2015Tri-axialLoops}. The recent development of the Bluetooth LE Audio standard with its ``Auracast'' public broadcast implementation that was published in July 2022 promises the same direct-to-hearing aid connectivity that induction loop systems provide. Auracast will also be compatible with cochlear implants and personal listening devices, such as headphones or earphones.

\section{Problem}

It is very likely that the increased signal-to-noise ratio provided by an ALS would be beneficial to almost everyone in almost any listening environment. In reality, however, ALSs are typically only used by individuals with severe to profound hearing losses. The reasons for this lack of mass adoption are poorly understood and, to date, have not been systematically investigated, as the field of hearing assistive technology has been historically underfunded and under-researched.

Hearing loss has been associated with a wide variety of physical and mental health risks, including social isolation, depression, lower quality of life and a significantly elevated risk of dementia \cite{atef_impact_2023, huang_hearing_2023, wen_factors_2023}. On average, individuals wait 9 years from hearing aid candidacy to hearing aid adoption, while earlier adoption may mitigate some of the risks associated with hearing loss \cite{knoetze_factors_2023, simpson_time_2019}. Research has shown that individuals who have access to assistive listening technology via a telecoil in their hearing device exhibit higher satisfaction with their devices overall \cite{picou_hearing_2022}. For these reasons, mass adoption of assistive listening technology would be a desirable outcome for individuals with hearing loss and society as a whole.

We hypothesize that the following five reasons may hinder mass adoption of assistive listening technology by the general public and propose potential solutions for each.

\subsection{Personally Owned Receiving Devices}

Listeners may prefer to use personally owned receiving devices over borrowing a receiving device from a venue. The reasons for this preference may include convenience, comfort, and hygiene. Given the variety of frequency bands used for RF and IR systems, it may be rare that listeners own a compatible receiving device. Audio frequency induction loop systems, however, use a globally universal standard for transmitting an audio signal to a listener. Therefore, the vast majority of hearing aid models and all cochlear implants in the market today are compatible with the technology. Auracast will also allow listeners to listen with personally owned receiving devices.

\subsection{High-Fidelity Stereo Sound}

Listeners may prefer to listen to stereo sound rather than to mono sound, particularly given that listening to stereo sound with earphones, headphones, hearing aids, or cochlear implants would allow for a listening experience with full spatialization, as demonstrated by Starkey's ``Virtual Barber Shop'' experience, which can easily be found online. Listeners may also prefer a high-fidelity listening experience, which may not always be available with a variety of assistive listening technologies due to low audio bandwidth, distortion, interference, delay, and packet loss.

High-fidelity stereo sound is not currently available with any of the legally compliant assistive listening technologies. However, high-fidelity stereo sound will be available with Auracast.

\subsection{Unoccluded Listening Experiences}

ALSs are mostly used in group settings where many other listeners are present in the audience. A listener may prefer to maintain situational awareness and the ability to converse with others in their proximity while listening to the presented audio signal. This would require an unoccluded listening experience, which is typically not available for venue-owned receiving devices with headphones. 

While listeners could provide their own open-back headphones or a bone conduction headset, such an occurrence may be rare. Individuals with hearing aids may have open-fit hearing aids, providing an unoccluded listening experience. For any listener with a hearing aid or cochlear implant, the equivalent of an unoccluded listening experience may be achieved via a program that mixes the assistive listening audio signal and the acoustic audio signal received by the device's microphone. This is similar to the ``transparency mode'' that any contemporary headphones and earphones offer. This mode allows listening to a streamed audio signal from a smartphone, for example, while feeding the signal received by the device's external microphones to the output transducer, creating the illusion of an unoccluded listening experience. Individuals with only one hearing device and a healthy other ear would also have access to a partially unoccluded listening experience.

 Auracast will be able to leverage the transparency mode of compatible receiving devices to create an unoccluded listening experience with high-fidelity stereo sound.

\subsection{Distortion from Alignment Delay}

The audio signal that is transmitted via a venue's ALS typically travels at the speed of light (299,792,458\,m/s). The acoustic signal produced by the venue's loudspeaker system travels at the speed of sound, which is approximately 343\,m/s or 1.125\,ft/ms in dry air at $20\,^\circ\mathrm{C}$. The result of this is that the assistive listening audio signal is perceived by the listener instantaneously, while the acoustic audio signal is perceived with a delay. The magnitude of the delay is dependent upon the distance between the listener and the venue's front loudspeakers, which are typically mounted in alignment with the front edge of the stage. For example, if the listener is seated at a distance of approximately 200\,ft / 60\,m from the stage, the acoustic signal from the loudspeakers arrives at the listener's ears approximately 180\,ms after its origination. This delay is not of concern when the acoustic signal is the only signal the listener perceives. However, if the listener perceives both the assistive listening audio signal and the acoustic signal at the same time and one of the two signals is delayed with respect to the other signal, the listener perceives a distorted signal.

For very short delay times of approximately 5\,ms or less, this distortion is perceived as a ``coloration'' of the sound due to comb-filter effects between the two signals \cite{Stiefenhofer2022HearingHearing-impaired}. For delay times longer than 5\,ms but shorter than 30\,ms, the distortion tends to be perceived as a reverberation, while delay times greater than 30\,ms tend to be perceived as an echo \cite{Litovsky1999TheEffect,Bramslow2010}. Such distortions reduce speech intelligibility and increase listening effort \cite{Lelic2022HearingHearing}. A multitude of experiments have found that the threshold at which most listeners consider the perceived echo effects intolerable is approximately 30\,ms, while individuals with hearing loss tend to tolerate slightly longer delay times than individuals with normative hearing \cite{Goehring2018TolerableAids}. Recent research has also shown that both listeners with and without hearing loss generally prefer shorter delay times \cite{Lelic2022HearingHearing}.

In an assistive listening situation, the perceived distortion effect is greatest for individuals who have an unoccluded listening experience as described above, for example as the result of using open-fit hearing aids. It is assumed that in today's market, more than 50\% of hearing aids are being dispensed with an open fit \cite{Froehlich2019ClosingUnderstanding.}. Given that the vast majority of ALS users in any given venue are perceptible to this delay effect and that the maximum tolerable alignment delay is approximately 30\,ms, any venue in which the distance between the stage edge and the furthest seat from the stage exceeds 35\,ft / 10\,m, should be equipped with an assistive listening solution that accounts and compensates for this delay effect.

None of the currently available assistive listening technologies have the ability to accurately account and compensate for alignment delay effects. Alignment delay compensation is a requirement for \emph{any} type of current or future assistive listening technology if an optimal listening experience is desired. Proposed solutions will be discussed below.

\subsection{Automatic Connectivity}

Listeners may prefer that their assistive listening receiving device connects automatically to an available assistive listening audio stream. This preference may be strongest for listeners who wear their hearing devices for the majority of their waking hours and who transition frequently between listening environments. These listeners may be users of hearing aids, cochlear implants, or earphones with transparency mode, for example. 

Currently, connecting to a RF or IR system requires the listener to borrow a receiving device from the venue. Connecting to an induction loop system requires users to press a button on their hearing device, on a remote control, or to select an induction loop program on their smartphone.

For Auracast, individuals may be exposed to a plurality of available broadcast audio streams, therefore creating ambiguity which audio stream to connect to, requiring the listener to manually connect to an audio stream, similarly to connecting to a public Wi-Fi network. Proposed methods to connect to an Auracast audio stream include selecting a stream on a smartphone or remote control device, tapping a smartphone against a near-field communication (NFC) device, or scanning a QR code. None of these methods provide the same level of accessibility as connecting to an induction loop system. This may pose a hurdle to individuals who, for whatever reason, may be unable to operate a smartphone or remote control device.

None of the currently available assistive listening technologies have the ability to automatically connect to an assistive listening audio stream. A proposed solution will be discussed below.

\section{Solutions}

Currently available assistive listening technologies exhibit a variety of shortcomings that may prevent their mass adoption by the general public and may hinder wider adoption of hearing aids. Of the technologies discussed above, Auracast holds the most promise for a globally universal standard that has the ability to meet all problems discussed above. However, to date, the requirement of compensating for alignment delay and the issue of automatic connectivity have not been solved. We propose potential solutions to both of these concerns.

\subsection{Distortion from Alignment Delay}

The requirement to compensate for the alignment delay that results from the slower travel speed of the acoustic audio signal in comparison to the assistive listening audio signal could be solved in a variety of ways.

\subsubsection{Multiple Broadcasts}

One possible solution may be to equip a venue with multiple assistive listening transmitters. Each of these transmitters would have to send an audio signal with an additional alignment delay, whereby the alignment delay would have to be adjusted to a specific audience segment. For example, a venue with a maximum distance of 200\,ft / 60\,m from the stage edge to the furthest seat, such as a large performing arts venue with more than 2,000 seats, would require 3 assistive listening transmitters to ensure that every seat experiences an alignment delay of less than ±30\,ms. With increasing venue size and/or stricter requirements for the maximum tolerable delay time, the number of required transmitters increases.

This approach is currently being followed for complex induction loop system installations. Multiple induction loop amplifiers are fed by a digital signal processor that adds an appropriate amount of alignment delay for each audience segment in which an array of antenna wires is installed. However, the provided alignment delay is only an approximation for each audience segment and is not accurate to each listener's position. The ``selection'' of the desired signal is inherent in the listener's physical position due to induction loop antennas being designed as near-field antennas. For RF and IR systems, however, listeners or venue staff would have to select the desired channel based on the listener's seat location. This solution poses additional challenges in venues with standing room as well as at outdoor music events, for example.

The current Bluetooth LE Audio specification defines a parameter with the label ``Presentation Delay'' that is broadcast by the transmitter (Broadcast Source) in extended advertising packets, together with metadata describing the parameters of the actual audio stream, including codec settings and channel information. While the primary purpose of the Presentation Delay parameter is to synchronize multiple receiving devices (Broadcast Sinks), it may also be used to provide appropriate alignment delay for audience segments.

However, for the Broadcast Sink to render a presentation delay, the Broadcast Sink needs to be able to buffer the audio data from the time it is received until it is presented to the listener. The current version of the Bluetooth LE Audio specification does not require support of Presentation Delay values of more than 40\,ms. Therefore, it may be possible that many Auracast-compatible receiving devices may not be able to provide enough buffer time to achieve temporal alignment between the received assistive listening audio signal and the acoustic audio signal in large venues.

Further, commissioning multiple broadcast streams with Auracast within current specifications  may result in a multitude of identical audio streams being broadcast, leading to air time and spectrum occupancy limitations. In Bluetooth LE Audio, the audio plane is separated from the control plane, meaning that broadcast audio streams only contain raw audio data packets, while command and control parameters are transmitted separately via advertising packets. Therefore, it may be possible to provide only one broadcast audio stream together with multiple extended advertising packet streams, each containing a unique value for Presentation Delay, and all of them pointing to the same broadcast audio stream. However, this configuration is not allowable within current Bluetooth LE Audio specifications. Current Bluetooth LE Audio specifications also do not provide a parameter to configure a local alignment delay that is specific to each listener.

\subsubsection{Variable Alignment Delay at Receiving Device}

Another solution to ensure temporal alignment between the assistive listening audio signal and the acoustic audio signal may be an assistive listening receiver with a ``delay compensation'' feature, which adds a variable amount of alignment delay after receiving an assistive listening audio signal from a transmitter and prior to presenting the audio signal to the listener. The amount of alignment delay needs to be customizable to each listener's specific distance from the stage or furthest loudspeaker array, ensuring perfect or near-perfect time-alignment, thus creating the best possible listening experience without compromises for all audience members.

\subsubsection{Ultrasound Transmitter}

A third option to automatically compensate for the experienced alignment delay may be an ALS that uses an ultrasound carrier wave. An ultrasound carrier travels at the speed of sound, just like the acoustic audio signal does. Therefore, the assistive listening audio signal and the acoustic audio signal would arrive at the listener's ear at the same time. Currently available MEMS microphones allow receiving ultrasound signals of up to 100\,kHz or more, providing ample bandwidth for a high-fidelity stereo audio signal. Many hearing aids and cochlear implants already contain such microphones.

It may also be possible to transmit data via this ultrasound carrier. This feature could be utilized to transmit a real-time transcription of the audio stream to the listener for display on a smartphone or tablet, for example.

\subsection{Automatic Connectivity}

To solve the requirement for automatic connectivity to an available assistive listening audio stream, a few assumptions need to be made. It can be assumed that a listener may be exposed to multiple broadcast streams simultaneously. It can further be assumed that when a assistive listening audio stream is provided, at least one of the broadcast streams will correlate with the listener's acoustic environment. It can lastly be assumed that the broadcast stream with the greatest correlation to the listener's acoustic environment is the one the listener would want to connect to.

Given these assumptions, the requirement for automatic connectivity can be solved by (1) scanning the environment for available broadcast audio streams, (2) comparing each of the streams to the listener's acoustic environment, and (3) automatically connecting to the audio stream that most closely resembles the listener's acoustic environment. This selection could be suppressed or overwritten by the listener via a push button, voice control, a smartphone, or a remote control, for example.

\section{DISCUSSION}

To achieve mass adoption of assistive listening technology, the resulting listening experience needs to meet requirements of audiophile listeners without hearing loss and provide a seamless and smooth user experience. We propose a variety of solutions to the shortcomings of currently available assistive listening technology.

To ensure success of \textit{any} type of ALS in venues where \textit{any} listener is seated at a distance greater than 35\,ft / 10\,m, the ALS needs to be equipped with a method to compensate for the alignment delay that results from the slower travel speed of the assistive listening audio signal compared to the acoustic audio signal from the venue's loudspeaker system. This could be achieved by providing multiple broadcast streams, equipping receiving devices with the ability to add a variable amount of alignment delay, or by utilizing an ultrasound carrier for the assistive listening audio signal.

Bluetooth Auracast promises to make assistive listening technology more widely available, accessible, and affordable, and may subsequently increase adoption of hearing aids. However, the only available method within the current Bluetooth LE Audio specification that allows to temporally align the assistive listening audio signal and the acoustic audio signal may lead to air time and spectrum occupancy limitations in large venues. The Bluetooth LE Audio specification should be amended to (1) allow for multiple extended advertising packet streams to point to the same broadcast audio stream, (2) require Broadcast Sinks to provide support for Presentation Delay values of at least 500\,ms, (3) provide a parameter for a remote control device (Commander / Broadcast Assistant) to communicate a ``Local Alignment Delay'' parameter to the Broadcast Sink, and (4) require that Broadcast Sinks provide the option to automatically connect to a broadcast stream that has a high correlation to the listener's acoustic environment.

\newpage
\bibliographystyle{IEEEbib}
\bibliography{references}

\end{document}